\documentclass[epj,final]{svjour}

\usepackage[dvips]{epsfig}
\usepackage[dvips]{graphicx}
\usepackage{latexsym,verbatim}
\usepackage{bm}
\usepackage{hyperref}

\newcommand{\av}[1]{\langle {#1} \rangle}

\begin{document}

\title{Heterogenous mean-field analysis of a generalized voter-like
  model on networks}

\author{Paolo Moretti\inst{1,2} \and Suyu Liu\inst{3} \and Andrea
Baronchelli\inst{1,4} \and Romualdo Pastor-Satorras\inst{1}}

\institute{Departament de F\'isica i Enginyeria Nuclear, Universitat
Polit\`ecnica de Catalunya, Campus Nord B4, 08034 Barcelona, Spain
\and Departamento de Electromagnetismo y F\'isica de la Materia, Facultad de Ciencias, Universidad de Granada, Fuentenueva s/n, 18071, Granada, Spain
\and State Key Lab. of Industrial Control Technology, Institute of
Cyber-systems and Control, Zhejiang University, Hangzhou 310027,
China
\and Department of Physics, College of Computer and Information Sciences,
Bouv\'e College of Health Sciences, Northeastern University, Boston,
MA02115, USA}
\date{\today} 

\abstract{We propose a generalized framework for the study of voter
  models in complex networks at the heterogeneous mean-field (HMF)
  level that (i) yields a unified picture for existing copy/invasion
  processes and (ii) allows for the introduction of further
  heterogeneity through degree-selectivity rules. In the context of
  the HMF approximation, our model is capable of providing
  straightforward estimates for central quantities such as the exit
  probability and the consensus/fixation time, based on the
  statistical properties of the complex network alone. The HMF
  approach has the advantage of being readily applicable also in those
  cases in which exact solutions are difficult to work out. Finally,
  the unified formalism allows one to understand previously proposed
  voter-like processes as simple limits of the generalized model.
  \PACS{ {PACS-key}{discribing text of that key} \and
    {89.75.-k}{Complex systems}\and {64.60.aq}{Networks}}
  }

\authorrunning{Paolo Moretti \textit{et al.}}
\titlerunning{Heterogenous mean-field analysis of a generalized voter-like
  model on networks}

\maketitle

\section{Introduction}

A topical problem in the statistical physics approach to social and
evolutionary dynamics \cite{Castellano09,DrosselEvolution} is the
study of the mechanisms ruling the formation of consensus in an
initially disordered population, in situations implying the opinion
about a certain issue, the intention of voting in an election, or the
evolutionary competition of different species striving for the same
ecological resources. Several stochastic copying/invasion processes
have been proposed to represent this kind of problems, the simplest
being the voter model \cite{Clifford73} and the Moran process
\cite{moran1958random}. In these models, each individual in a
population (agent) is endowed with a binary variable (opinion or
state) with value $\sigma=\pm1$. At each time step, an agent $i$,
together with one nearest neighbor $j$, are selected at random. In the
voter model the system is updated as $\sigma_i := \sigma_j$, the first
agent copying the opinion of its neighbor. The Moran process, on the
other hand, can be considered as a \emph{reversed} voter model, in
which it is the neighboring agent the one who copies the opinion of
the first agent, $\sigma_j := \sigma_i$. Starting from a disordered
initial state, this kind of dynamics leads in finite systems to a
uniform state with all individuals sharing the same state (the
so-called consensus). The way in which this final state is reached is
usually characterized in terms of the exit probability $E(x)$ and the
consensus time $T_N(x)$, defined as the probability that the final
state corresponds to all agents in the state $+1$ and the average time
needed to reach consensus in a system of size $N$, respectively, when
starting from a homogeneous initial condition with a fraction $x$ of
agents in state $+1$ \cite{Castellano09}.

While simple voter-like models are well understood on regular
lattices, even in terms of theorems and exact solutions
\cite{liggett1985ips,KineticViewRedner}, they become more relevant in
social and evolutionary contexts when considered on top of complex
networks, which act as more realistic representations of social or
ecological contact patterns
\cite{barabasi02,mendesbook,newman2003saf}. The analysis of the voter
model in these substrates reveals nontrivial differences with respect
to ordered lattices. For example, now the order in which interacting
individuals are selected becomes relevant \cite{Suchecki:2005p107}, in
such a way that voter model and Moran process behave in different
ways. Moreover, relevant quantities such as the consensus time turn
out to depend on the heterogeneity of the contact pattern, as measured
by the degree distribution
\cite{PhysRevLett.94.178701,Antal06,Sood08}. Not only the properties
of the interaction substrate are relevant in this case, but also the
intrinsic heterogeneity of the actors. Their individual propensity to
interact with peers, and change state accordingly, plays a significant
role \cite{schneider09,Yang09,Lin10,2010arXiv1011.2395B}. Other
sources of heterogeneity that have been considered include explicitly
the directed nature of the connections
\cite{1742-5468-2009-10-P10024}.

The theoretical understanding of voter-like models (and dynamical
processes in general) on complex networks has been traditionally
accomplished by application of heterogeneous mean-field (HMF)
approaches \cite{barratbook,dorogovtsev07:_critic_phenom}, which are
based on a twofold assumption: (\emph{i}) The network description is
coarse-grained into degree classes, all vertices in the same class
having the same degree and sharing the same dynamical properties;
(\emph{ii}) The real (\textit{quenched}) network structure is replaced
by an \textit{annealed} one \cite{dorogovtsev07:_critic_phenom}, which
disregards the actual connection pattern and simply assumes that the
degree class $k$ is connected to the degree class $k'$ with
conditional probability $P(k'|k)$ \cite{marian1}. Very significant
progress has been achieved in the analysis of voter-like models within
the HMF approach, which allows one to work out simple analytic expressions
for the quantities of interest, showing reasonable agreement with
numerical simulations in real quenched networks
\cite{PhysRevLett.94.178701,Antal06,Sood08,schneider09,2010arXiv1011.2395B,1742-5468-2009-11-L11001}.

Recently, a generalized formalism for the class of heterogeneous
stochastic-copying voter-like models on networks has been proposed
\cite{baxter:258701}, in which the process is identified by the
copying rate $C_{ij}$, encoding the full structure of the contact
network and the stochastic update rules, and defined as the rate at
which vertex $i$ in the network copies the state of vertex $j$. Thus,
for example, the standard voter model corresponds to the choice
$C_{ij} = a_{ij}/[N k_i]$ with $k_i=\sum_{r}a_{ir} $, where $a_{ij}$
is the adjacency matrix of the network and $N$ the network size. The
Moran process, analogously, is given by $C_{ij} = a_{ij}/[N k_j]$.
Within this formalism, it has been shown that both the exit
probability and the consensus time can be calculated exactly from the
knowledge of the spectral properties of the matrix $C_{ij}$, provided
that certain general conditions are met
\cite{baxter:258701,1751-8121-43-38-385003}, a result that lays the
foundation for a mathematical understanding of general copying
processes and their mapping to particle-reaction systems. Despite the
fact that the formalism in
Refs.~\cite{baxter:258701,blythe07:_stoch_model} is exact, and
provides in some cases more accurate results than HMF theory, it is
still useful to consider general stochastic-copying models from the
perspective of HMF theory. In this framework, indeed, approximate
analytical results can be obtained when the exact solution would be
hard to work out in practice. For example, in realistic heterogeneous
environments involving large numbers of agents, explicit expressions
for $C_{ij}$ might not be readily accessible. Moreover, the spectral
properties of the copying rates are in general non-trivial to obtain,
unless the matrix $C_{ij}$ has a relatively simple form.

In this paper we pursue this path, proposing a generalized
coarse-grained voter-like model on networks and showing how the HMF
approach allows us to obtain very simple estimates for central
properties such as the exitœ probability and the consensus time. We
check the validity of our approach by considering a simple example of
opinion dynamics in a homophilic society, in which vertices with
similar degree are more prone to interact than vertices with differing
degree.

\section{Generalized voter model on networks} 
\label{sec:generalized_voter_model_on_networks}


Inspired by Ref.~\cite{baxter:258701}, we consider a stochastic 
model on networks defined in terms of a heterogeneous voter model as
follows: 
\begin{itemize}
\item Each vertex $i$ is endowed with a given fitness $f_i$
  \cite{Antal06}.
\item A source vertex $i$ is selected at random, with a probability
  $f_i / \sum_j f_j$, i.e., proportional to its fitness $f_i$.
\item A nearest neighbor $j$ of $i$ is then selected at random.
\item With probability $Q_{ij}$, $i$ copies the state of vertex
  $j$. Otherwise, nothing happens.
\end{itemize}
With these settings, the microscopic copying rate $C_{ij}$, as
considered in Ref.~\cite{baxter:258701} will be given by
\begin{equation}
	C_{ij} = \frac{f_i }{\sum_j f_j} \frac{a_{ij}}{k_i} Q_{ij}.
\end{equation}

In the spirit of the HMF approximation, we can replace the microscopic
copying rate by its degree class average.  The quantities $f_i$ and
$Q_{ij}$ are simply coarse-grained by averaging them over the set of
vertices with a given fixed degree, i.e.
\begin{eqnarray}
  f_i  &\to& \frac{1}{N P(k)} \sum_{i \in k} f_i \equiv f_k,\\
  Q_{ij} &\to& \frac{1}{N P(k)}\frac{1}{N P(k')} \sum_{i \in k}
  \sum_{i \in k'}  Q_{ij} \equiv Q(k,k'), 
\end{eqnarray}
where $i \in k$ denotes a sum over the degree class $k$ and $P(k)$ is
the network's degree distribution. For the term concerning the random
choice of a nearest neighbor, we follow
Ref.~\cite{dynam_in_weigh_networ} to substitute
\begin{equation}
  \frac{a_{ij}}{k_i} \to  \frac{[N P(k)]^{-1} \sum_{i \in k} \sum_{j
      \in k'}  
    a_{ij}}{[N P(k)]^{-1}  \sum_{i \in k} \sum_{r} a_{ir}} \equiv
  P(k'|k). 
\end{equation}
At the coarse-grained degree level, our generalized voter model is
thus defined in terms of the mesoscopic copying rate
\begin{equation}
  C(k, k')\equiv\frac{f(k)}{\langle f(k) \rangle} P(k'|k) Q(k,k'),
\end{equation}
where the function $f(k)$ comes with its proper normalization factor.

\section{Heterogeneous mean-field solution} 
\label{sec:heterogeneous_mean_field_solution}


In the HMF approach, ordering processes are quantified by studying the
evolution of the density of vertices of degree $k$ in the state $+1$,
$x_k$. In order to determine the rate equation satisfied by these
quantities \cite{PhysRevLett.94.178701,Sood08,2010arXiv1011.2395B}, we
consider the probability $\Pi(k; \sigma)$ that a spin in state
$\sigma$ at a vertex of degree $k$ flips its value to $-\sigma$ in a
microscopic time step. From the definition of the generalized voter
model, this probabilities can be simply written as
\[
\Pi(k; \sigma) = \left[\frac{1-\sigma}{2} + \sigma x_k\right]
\sum_{k'} \left[\frac{1+\sigma}{2} - \sigma x_{k'}\right] P(k)
C(k,k').
\]
From the previous expression, the rate equation for $x_k$ can be
written as \cite{2010arXiv1011.2395B}
\begin{eqnarray}
  \dot{x}_k &=&\frac{\Pi(k; -1)-\Pi(k; +1)}{P(k)} 
	\equiv \sum_{k'}C (k,k') (x_{k'}-x_k).\label{eq:first}
\end{eqnarray}

The model just posed is still too hard to solve even in the HMF
approximation. In the following, we will therefore make two major
simplifying assumptions leading to analytically solvable HMF
equations: (\emph{i}) Dynamics proceed on uncorrelated networks,
i.e. \cite{mendesbook}
\begin{equation}
	P(k'|k)=\frac{k'P(k')}{\langle k \rangle};
\end{equation}
and (\emph{ii})
the interaction probability can be factorized as
\begin{equation}\label{eq:voterQ}
  Q(k,k')=a(k)b(k')s(k,k'),
\end{equation}
where $s(k,k')$ is any symmetric function of $k$ and $k'$. As we will
see, this simplified form, which permits an analytic HMF solution,
still allows for a vastly rich phenomenology, encompassing all
voter-like models on complex networks previously proposed. Under this
conditions, defining
\begin{equation}
  u(k)=\frac{a(k)f(k)}{\langle f(k)\rangle}, \qquad
  v(k')= \frac{b(k')k'}{\langle k \rangle}, \label{eq:1}
\end{equation}
the HMF rate equation can be written as
\begin{equation}\label{eq:generalized}
\dot{x}_k=\sum_{k'}P(k')\Gamma (k,k') (x_{k'}-x_k)
\end{equation}
where $\Gamma(k,k')= u(k) v(k') s(k,k')$,

The HMF analysis  proceeds by first determining
the corresponding conservation laws
\cite{PhysRevLett.94.178701,Antal06,Sood08}. Conserved quantities for
the generalized process in Eq.~(\ref{eq:generalized}) can be
calculated as follows: We define a generic integral of motion
$\omega[x_k(t)]$ such that $d\omega/dt=0$. By definition of time
derivative, we have
\begin{equation}
  \frac{d\omega}{dt}=\nabla_{\bf x}\omega \cdot {\bf \dot{x}}=\sum_k
  \frac{\partial \omega}{\partial x_k} \dot{x}_k=0. 
\end{equation}
In analogy with previous results
\cite{PhysRevLett.94.178701,Antal06,Sood08}, we look for conserved
quantities that are linear in $x_k$ imposing $\partial \omega/\partial
x_k=z_k$ independent of $x_k$, so that conserved quantities will be
given by
\begin{equation}
\omega={\bf z} \cdot {\bf x} =\sum_k z_k x_k,
\end{equation}
where $z_k$ is any solution of $\sum_k z_k \dot{x}_k=0$ and
$\dot{x}_k$ is given by Eq. (\ref{eq:generalized}).  Considering the
explicit form of Eq. (\ref{eq:generalized}), the choice $z_k\propto
P(k) v(k)/u(k)$ always satisfies the above condition, so that a
conserved quantity is found up to multiplicative factors and additive
constants.  We choose the normalization $\sum_k z_k =1$, 
such that the conserved quantity is defined as
\begin{equation}
  \omega={\bf z} \cdot {\bf x} =\frac{\langle v(k)/
    u(k)\,x_k\rangle}{\langle v(k)/u(k)\rangle}.
\end{equation}
As for the usual voter model~\cite{PhysRevLett.94.178701} the
conservation law allows the immediate determination of the exit probability $E$, i.e. the probability that the final
state corresponds to all spins in the state $+1$.  In the final state
with all $+1$ spins we have $\omega=1$, while $\omega=0$ is the other
possible final state (all $-1$ spins).  Conservation of $\omega$
implies then $\omega = E \cdot 1+ [1-E] \cdot 0$, hence
\begin{equation}
  E = \omega = \frac{\langle v(k)/
    u(k)\,x_k\rangle}{\langle v(k)/u(k)\rangle}.
  \label{eq:3}
\end{equation}
Starting from a homogeneous initial condition, with a given density
$x$ of randomly chosen vertices in the state $+1$, we obtain, since
$\omega = x$,
\begin{equation}
  E_h(x) = x,
\end{equation}
completely independent of the defining functions $a$, $b$, and $s$,
and taking the same form as the standard voter model
\cite{Castellano09}. On the other hand, with initial conditions
consisting of a single $+1$ spin in a vertex of degree $k$, we have
\begin{equation}
  E_1(k) = \frac{v(k)/ u(k)}{N \langle v(k)/u(k)\rangle},
\end{equation}
which does not depend on the functional form of the symmetric
interaction term $s(k,k')$.

By looking at Eq. (\ref{eq:generalized}), every choice of $x_k$
constant in $k$ is a solution to the steady state condition
$\dot{x}_k=0$. We can prove that this solution is unique and does not
depend on initial conditions if the square matrix $P(k')\Gamma(k,k')$
is irreducible and primitive (it certainly is when working with
positive rates, which we will do in the following) \cite{gantmacher}.
We shall call the solution for the steady state $x_k(t\to
\infty)=x^\infty$. Then it is easy to prove that
\begin{equation}
  \omega= \sum_{k'} z_{k'} x_{k'}=x^\infty.
\end{equation}
that is, even in this general case, the steady state value for $x_k$
equals the conserved quantity.  

After a rapid exponential convergence to the steady state
distribution, the systems starts fluctuating diffusively around this
value, until consensus is reached. Such fluctuations characterize
finite systems and occur at long time scales, making such two-step
relaxation process possible in most cases.  The average consensus time
$T_N({\bf x})$ for a system in a generic state ${\bf x}$ can be
derived extending the well known recursive method to our general case
\cite{Sood08}.  At a given time $t$, $T_N({\bf x})$ must equal the
average consensus time at time $t+\Delta t$ plus the elapsed time
$\Delta t=1/N$ that is, in our notation,
\begin{equation}
  T_N({\bf x})=\bar\Pi \,T_N({\bf x})+\sum_{k,s}\Pi(k;s)T_N({\bf x}+{\bf \Delta x}^{(k)})
  +\Delta t,
\end{equation}
where $\bar\Pi=1-\sum_{k,s}\Pi(k;s)$ is the probability that no state
change occurs, while the sum is the weighted average over possible
state-updates ${\bf x}\to {\bf x}+{\bf \Delta x}^{(k)}$. The variation
${\bf \Delta x}^{(k)}$ is a vector whose all components are zero
except for the $k$-th, which equals the update-unit
$\Delta_k=[NP(k)]^{-1}$. Expanding to second order in $\Delta_{k}$,
taking $x_k=\omega$ as the initial state and changing variables such
that $\partial/\partial x_k=z_k\partial/\partial\omega$ we obtain the
backward Kolmogorov equation
\begin{equation}
-1=\frac{ {\bf z}^T\Gamma{\bf z} }{N} \omega(1-\omega)\frac{\partial^2T_N}{
\partial \omega^2}
\end{equation}
leading to
\begin{equation}\label{eq:TN}
  T_N(\omega)=-N_{\mathrm{eff}} [\omega \ln \omega + (1-\omega) \ln (1-\omega)]
\end{equation}
where we have defined the effective system size 
$N_{\mathrm{eff}} = N / \sum_{k,k'}z_{k}\Gamma(k,k')z_{k'}$,
which, in the case of generalized voter dynamics, 
Eq.~(\ref{eq:1}), becomes
\begin{equation}
  N_{\mathrm{eff}}=N
  \frac
  {
    \langle f(k)\rangle \langle k \rangle
    \left\langle\frac{kb(k)}{f(k)a(k)}\right\rangle^2 
  }
  {
    \left\langle\left\langle s(k,k') k
        b(k)\frac{[k'b(k')]^2}{f(k')a(k')}\right\rangle\right\rangle},
  \label{eq:2} 
\end{equation}
with
$\langle\langle\cdot\rangle\rangle=\sum_{kk'}P(k)P(k')(\cdot)$.

Equations~(\ref{eq:TN}) and (\ref{eq:2}), together with the expression
for the exit probability, Eq.~(\ref{eq:3}), represent the final HMF
solution of the generalized voter model. From these formulas it is
easy to recover most of the variations of the voter model considered
in the past. For example, the standard voter model is obviously
recovered for $a(k)=b(k)=f(k)=s(k,k')=1$. The invasion process
\cite{castellano05:_effec,Antal06}, also known as the Moran process in
the evolutionary literature \cite{moran1958random,nowak2006ed},
corresponds to $a(k)=k$, $b(k)=k^{-1}$ and $f(k)=s(k,k')=1$. Link
update dynamics \cite{Suchecki:2005p107} is recovered for
$a(k)=b(k)=s(k,k')=1$ and $f(k)=k$. The voter and Moran processes on
weighted networks characterized by a symmetric weight between vertices
of degree $k$ and $k'$ proportional to $g_s(k) g_s(k')$
\cite{2010arXiv1011.2395B} are reproduced by imposing $s(k,k')=1$ and
setting $a(k) = f(k)=1$, $b(k) = g_s(k) \av{k}/\av{kg_s(k)}$ and
$f(k)=1$, $a(k)=k g_s(k)/\av{kg_s(k)}$, $b(k) = \av{k}/k$,
respectively. Finally, an HMF implementation of the generalized voter
dynamics proposed in \cite{schneider09} is recovered imposing
$f(k)=a(k)=1$, $b(k)=k^{\alpha-1}$ and $s(k,k')=(k+k')/(k^\alpha+
k'^\alpha)$. A summary of the above mappings is presented in
Table~\ref{tab:summary}.

\begin{table}
  \begin{center}
    \begin{tabular}{c|cccc}
      Model & $f(k)$ & $a(k)$ & $b(k)$ & $s(k,k')$  \\  \hline
      Voter model \cite{Clifford73} & 1 & 1 & 1 & 1 \\
      Moran process \cite{moran1958random} & 1 & $k$ & $k^{-1}$ & 1\\
      Link update \cite{Suchecki:2005p107} & $k$  & 1 & 1 & 1 \\
      Voter weighted \cite{2010arXiv1011.2395B} & 1 & 1 &  $\frac{g_s(k)
        \av{k}}{\av{kg_s(k)}}$ & 1 \\ 
      Moran weighted \cite{2010arXiv1011.2395B} & 1  & $\frac{k
        g_s(k)}{\av{kg_s(k)}}$ & $\frac{\av{k}}{k}$ & 1 \\ 
      Generalized Voter \cite{schneider09} & 1 & 1 & $k^{\alpha-1}$ & $\frac{k+k'}{k^\alpha+
        k'^\alpha}$ \\ \hline
    \end{tabular}
  \end{center}
  \caption{Summary of the mapping of voter-like models to the present
    formalism.}
  \label{tab:summary}
\end{table}

\section{Numerical analysis} 
\label{sec:numerical_analysis}


In order to show an application of our formalism, we examine a toy
model to study the effects that homophily in social networks might
have in opinion formation dynamics. Homophily is broadly defined by
the tendency of people to interact with similar people
\cite{citeulike:3022278}. In the absence of information beyond the
topological structure of the contact network, the simplest assumption
 we can make is that homophily is driven by an increased tendency of
individuals to copy other individuals with a degree that is not too
different from their own. As a simple representation of homophilic
behavior we consider the case $s(k,k')=\exp[-R(k-k')/\xi^2]$ and
$f(k)=a(k)=b(k)=1$. Here $R(x)$ is any continuous even function with
$R(0)=0$ and a minimum in $x=0$, i.e. $R'(0)=0$ and $R''(0)>0$. The
parameter $\xi$ measures the amplitude of stochastic fluctuations
around ideal homophily: For $\xi\approx 1$ the probability of copying
neighbors with different degrees is strongly suppressed; for $\xi\gg
1$ fluctuations take over and simple voter behavior is rapidly
recovered.

In spite of its apparent
simplicity, this problem would be impossible to solve in a realistic
network by standard techniques \cite{baxter:258701}. By applying the HMF result in Eq.
(\ref{eq:2}) instead, we readily find
\begin{equation}\label{eq:sums}
  N_{\mathrm{eff}}^{\xi}=N\frac{
    \left[
      \sum_{k} kP(k) 
    \right]^3
  }
  {
    \sum_{k}\sum_{k'} \,
    kP(k)k'^2P(k')\exp\left[-\frac{R(k-k')}{\xi^2}\right]
  },
\end{equation}
where we  consider complex networks with a scale-free degree distribution
$P(k)\sim k^{-\gamma}$, with $\gamma \in ]2,3]$, minimum degree $m$
and maximum degree $k_c$.  In the limit of small $\xi$ the denominator
of Eq. (\ref{eq:sums}) can be evaluated in the continuous-degree approximation. 
By applying the Laplace
method for the first term in the asymptotic expansion, after an
adequate change of variables, it is thus straightforward to find that for small $\xi$
\begin{eqnarray}
  N_{\mathrm{eff}}^\xi&\simeq&
  \frac{N}{\xi}\left[\frac{R''(0)}{2\pi}\right]^{\frac{1}{2}}
  \times\nonumber\\
  &\times& \frac{2(1-\gamma)}{(2-\gamma)^2}\,
  \frac{(k_c^{2-\gamma}-m^{2-\gamma})^2} 
  {(k_c^{1-\gamma}-m^{1-\gamma})(k_c^{2-\gamma}+m^{2-\gamma})},
\end{eqnarray}
whereas in the opposite limit $\xi\to\infty$ one recovers the simple
voter model result, which in our notations reads
$N_{\mathrm{eff}}^\infty \sim N k_c^{\gamma-3}$.  
As a consequence,
the consensus time $T_N\propto N_{\mathrm{eff}}^{\xi}$ diverges for
small $\xi$ as $\xi^{-1}$ as the selectivity amplitude $\xi$
approaches zero. For increasing $\xi$, instead, $T_N$ decreases and
asymptotically crosses over to a plateau, where simple voter behavior
is recovered.  

\begin{figure}[t]
   \begin{center}
     \includegraphics*[width=8.4cm]{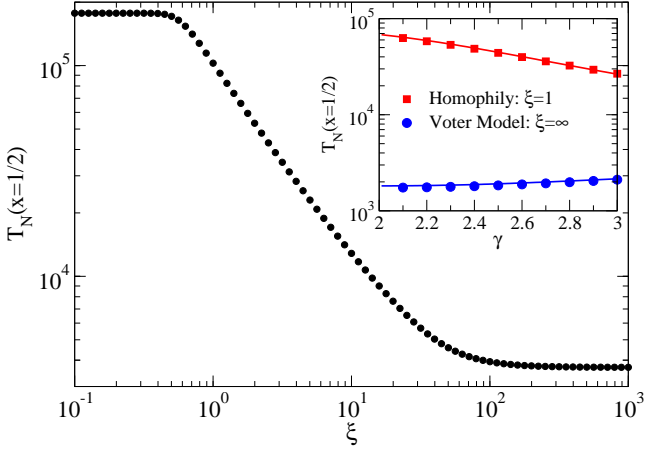}
     \caption{Consensus time as a function of the selectivity range
        from numerical evaluation of Eq. (\ref{eq:sums}). 
       The case of homogeneous initial conditions $x_k(0)=1/2$
       (maximum entropy) is considered. Inset: Consensus time as a function
       of the degree distribution exponent. Numerical estimates (points) and analytical 
       predictions (lines) based on $N_{\mathrm{eff}}^\infty$ and $N_{\mathrm{eff}}^\xi$ from the main text are compared.}
     \label{fig:Numerical}
   \end{center}
\end{figure}

\begin{figure}[t]
  \begin{center}
    \includegraphics[width=8.3cm]{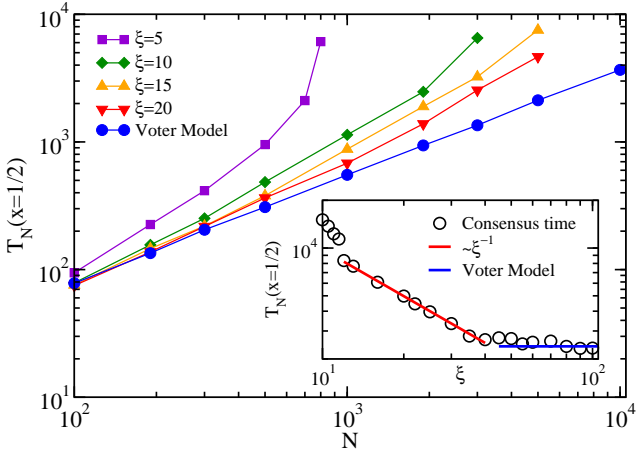}
    \caption{Consensus time with a Gaussian homophily factor in
      scale-free networks with $\gamma=2.5$, $m=4$ and $k_c\sim
      N^{1/2}$.  Main plot: Consensus time as a function of $N$, for
      different values of $\xi$. Inset: Consensus time at constant
      $N=5000$ as a function of $\xi$. The full line is a guide to the
      eye with slope $-1$.}
    \label{fig:Simulations}
  \end{center}
\end{figure}

In order to check explicitly the predictions of our formalism, we consider
a Gaussian homophily model, given by the simplest choice $R(x) = x^2$.
In Fig.~\ref{fig:Simulations} we plot the consensus times for
homogeneous initial conditions as a function of
$\xi$, computed from the numerical evaluation of
Eq.~(\ref{eq:sums}). We observe that, starting 
from large values of $\xi$,  $T_N$ is constant, as expected for simple 
voter behavior. Upon decreasing the selectivity amplitude $\xi$, the consensus time starts 
increasing, asymptotically behaving as $\xi^{-1}$. As soon as $\xi$ decreases below 
one, the discrete nature of the degree distribution takes over and $T_N$ reaches a plateau.
Interestingly, we find that $T_N$ is an increasing
function of $\gamma$ for large $\xi$, while it decreases with $\gamma$
in the small $\xi$ limit (Fig. \ref{fig:Numerical}, inset).

These observations have also been checked against direct numerical
simulations of the Gaussian ho\-mo\-phi\-ly mo\-del on uncorrelated
scale-free networks, generated with the uncorrelated configuration
model \cite{ucmmodel}. In Fig.~\ref{fig:Simulations} we show (main
plot) the consensus time as a function of the network size $N$ for
different values of $\xi$. We can see that the overall plot increases
at fixed $N$ for decreasing values of $\xi$, while for large $\xi$
it tends to the limit given by the simple voter model. When
considering $T_N$ as a function of $\xi$ in a network of fixed size,
inset of Fig.~\ref{fig:Simulations}, we observe, as expected, an
increase for decreasing $\xi$, followed by a plateau for large $\xi$.
At intermediate values of $\xi$, a reasonable fit to the form
$\xi^{-1}$ can be obtained. For smaller values of $\xi$, the selectivity 
range is so narrow that quenched effects of the network topology take over,
tending to slow down the
dynamics beyond the HMF prediction \cite{2010arXiv1011.2395B}.
Values of $T_N$ fluctuate wildly and deviate from the predictions for
annealed-network topologies.

\section{Conclusion}

In this paper we have studied a generalized voter-like model in the
framework of the HMF theory. Remarkably, the HMF formalism we have
adopted allows for a straightforward mapping of most of the voter-like
models proposed in the past. Not only the the properly said voter
model, the Moran process, and their weighted generalization can be
reproduced in our formalism, but also the link-update dynamics and
recently introduced generalizations of the voter model can be
straightforwardly recovered. The HMF approach allows for predictions
that are in fair agreement with the exact results whenever they are
available, and most importantly provides approximate solutions in the
event that exact methods are not viable.

As an instance of the versatility of the HMF approach, we have
considered a simple example of a heterogeneous voter model in which
vertices with similar degree tend to interact more often among
themselves rather than with the rest of the network. This degree
selectivity is a natural HMF implementation of the concept of
homophily, i.e. the tendency of individuals to associate and bond with
similar others. While an exact solution to such problem would be
unworkable, we have shown that in this framework homophily ends to
slow down consensus dynamics, in reasonable agreement with the
predictions of HMF theory.

In the future, it would be interesting to extend the approach to
voter-like methods that exhibit a surface tension, as for example the
Naming Game \cite{Baronchelli_2006}, or the noise-reduced voter model
\cite{dallasta_2007}, as well as to improve the quality of our present
results by applying more sophisticated HMF approaches, such as the
dynamical pair approximation \cite{1367-2630-10-6-063011}, or the
master equation approach \cite{gleeson2011high}.

\begin{acknowledgement}
  P.M., A.B., and R.P.-S.  acknowledge financial support from the
  Spanish MEC, under project FIS2010-21781-C02-01, and the Junta de
  Andaluc\'{\i}a, under project No. P09-FQM4682. A.B. acknowledges
  support from the Spanish Ministerio de Ciencia e Innovaci\'{o}n
  through the Juan de la Cierva program. R.P.-S. acknowledges
  additional support through ICREA Academia, funded by the Generalitat
  de Catalunya. S.L. acknowledges financial support from National Natural 
  Science Foundation of China (60421002).
\end{acknowledgement}

\bibliographystyle{epj}

\end{document}